\documentclass[11pt,twoside]{article}
\usepackage[dvips]{graphics,color}
\usepackage{array}
\newcommand{\be}{\begin{equation}}
\newcommand{\ee}{\end{equation}}
\newcommand{\ben}{\begin{eqnarray}}
\newcommand{\een}{\end{eqnarray}}
\newcommand{\benn}{\begin{eqnarray*}}
\newcommand{\eenn}{\end{eqnarray*}}
\newcommand{\KUCPlogo}{\hbox{\lower 1.4ex\hbox{\Huge\boldmath $\cal K$}
\kern -1.15em {\sffamily \bfseries\large\ UCP}
\put(-27.5,-6){\tiny\it preprint}
}}
\begin{document}
\begin{titlepage}
\begin{flushleft}
\KUCPlogo
\end{flushleft}
\vspace{-20mm}
\begin{flushright}
KUCP0173
\end{flushright}
\vspace{30mm}
\centerline{\LARGE Universal Structure of the Personal Income Distribution}
\vspace{10mm}
\centerline{\large Wataru Souma}
\vspace{5mm}
\centerline{\large souma@phys.h.kyoto-u.ac.jp}
\vspace{10mm}
\centerline{\large\it Faculty of Integrated Human Studies,}
\vspace{0mm}
\centerline{\large\it Kyoto University,
Kyoto 606-8501, Japan}
\vspace{50mm}
\begin{abstract}
We investigate the Japanese personal income distribution
in the high income range over the 112 years 1887-1998, and
that in the middle income range over the 44 years 1955-98. 
It is observed that the distribution pattern of the lognormal
with power law tail is the universal structure.
However the indexes specifying the distribution differ from
year to year.
One of the index characterizing the distribution is the mean value of the
lognormal distribution; the mean income in the middle income range.
It is found that this value correlates linearly with the Gross Domestic
Product (GDP).
To clarify the temporal change of the equality or inequality of
the distribution, we
analyze Pareto and Gibrat indexes, which
characterize the distribution in the high income range and
that in the middle income range respectively.
It is found for some years that there is no correlation
between the high income and the middle income.
It is also shown that the mean value of Pareto index equals to $2$, and
the change of this index is effected by the change of the asset price.
From these analysis we derive four constraints that must be satisfied by
mathematical models.
\end{abstract}
\end{titlepage}
\section{Introduction}
The personal income is a basic ingredient
of the economics.
Hence, a study of its distribution and
an investigation of the mechanism behind it 
have important meaning in the econophysics \cite{econo}.

The study of the personal income has long history
and many investigations have been done.
The starting point is about one hundred years ago when
V.~Pareto proposed the power law distribution of the personal income
\cite{Pareto}.
He analyzed the distribution of the personal income
for some countries and years, and found that
the probability density function $p(x)$ of the personal income $x$
is given by
\[
p(x)=Ax^{-(1+\alpha)}.
\]
Here $A$ is the normalization constant.
This power law behavior is called Pareto law and
the exponent $\alpha$ is named Pareto index.
This is a classic example of fractal distributions \cite{Mandelbrot}.
If Pareto index has small values, the personal income is unevenly distributed.

However, today, it is well known
that Pareto law is only applicable to
the high income range.
It was clarified by R.~Gibrat that the distribution takes the form of the
lognormal distribution in the middle income range \cite{Gibrat}.
As is well known the probability density
function in this case is given by
\[
p(x)=\frac{1}{x\sqrt{2\pi\sigma^2}}\exp
\left[-\frac{\log^2\left(x/x_0\right)}{2\sigma^2}\right],
\]
where $x_0$ is a mean value, and $\sigma^2$ is a variance.
Sometimes $\beta=1/\sqrt{2\sigma^2}$ is called Gibrat index.
The large variance means the global distribution of the income.
Hence the small $\beta$ corresponds to the uneven distribution of the
personal income.

The lognormal distribution with power law tail
for the personal income is rediscovered in the beginning of the 80s.
In Ref.~\cite{lognormal}, the investigation were performed for
the USA's personal income data for the fiscal year 1935-36.
It was confirmed that
the top 1\% of income follows Pareto law with $\alpha=1.36$
and the other takes the form of the lognormal distribution.
The explanation of the distribution pattern is
performed from a thermodynamical point of view.

From the age of J.J.~Rousseau,
one of the subject
of the social science is the theory of the inequality.
In economics, many indexes specifying the
unevenness of the income distribution have been developed.
Among them,
Gini index is well known and used
in many aspects of the economics.
Although Gini index is useful for the measure of the unevenness
in the economics,
this index has no attraction from the physical point of view.
This is because, as shown in this article,
different mechanisms are working in each regions of
the high income and the middle income.

Although many investigations of the personal income distribution
have been performed, data sets are all old.
Hence to reanalyze the income distribution by the recently
high quality data is meaningful.
In a previous article \cite{Aoyama}, we analyzed two data sets to gain the
personal income distribution of Japan in the year 1998.
One is the coarsely tabulated data containing all 6,224,254 workers
who filed tax returns.
The other is the list for the income-tax of 84,515 workers. 
In that article
we translate income-tax data to the income data
and connected these two data sets under some assumptions.
From these analysis, for the highest income region, we get the result that
the personal income distribution of annual year 1998 in Japan
follows Pareto law with $\alpha=2.06$.
However, this analysis is incomplete
for the middle income region, because
the data set for this region is sparse.
Hence, the first purpose of this article
is to gain the overall profile of the personal income distribution
of Japan in the year 1998.
For this purpose, we add sample survey data for the employment income. 
These are discussed in section 2.

The second purpose is focused on the change of Pareto index $\alpha$,
the mean value of the lognormal distribution $x_0$,
and Gibrat index $\beta$.
Although these indexes have been estimated in many countries and many years,
a study of the succeeding changes of these indexes is
not well known. Hence the investigations of the temporal change
of these indexes have important meaning.
In addition, we study the correlation between $\alpha$ and $\beta$.
This is because,
if the change of the distribution of the high income
and that of the middle income consists in the same mechanism,
the explicit relation between them is expected.
Moreover to study the reason of the change of $\alpha$ more detail,
we investigate the correlation between $\alpha$
and asset prices, especially the land price index
and the Tokyo Stock Price Index (TOPIX).
These are discussed in section 3.
Section 4 is devoted to the summary and discussion.
\section{Lognormal distribution with power law tail}

The {\itshape income data} contains the person who filed tax return individually.
This is a coarsely tabulated data.
We analyze this data over the 112 years 1887-1998 in this article.
The data is publicly available from the Japanese Tax Administration (JTA)
report
and the recent record is on the web pages of the JTA \cite{JTA1}. 
The example for the year 1998 is listed in the second column in
Table. The first column of Table is the
classification of the income $x$ with the unit of million yen.

The {\itshape employment income data} is the 
sample survey for the salary persons working in the private enterprises,
and does not contain the public servants and the persons with
daily wages.
This data is coarsely tabulated as same as the income data.
We analyze this data over the 44 years 1955-98 in this article.
This is publicly available from the JTA report and
recent record is available on the same web pages for the income data.
The distribution is recorded by the unit
of thousand people from the year 1964, therefore
the data contains the effect of the round-off.
The example of this data for the year 1998 is listed in the third
column in Table.

The {\itshape income-tax data} is only
available for the year 1998.
This is the list of the 84,515 individuals who paid the income-tax of
ten million yen or more in that year.
The contained data is reported by the JTA.

To gain the overall profile of the personal income distribution,
we connect these data sets.
As mentioned in the previous article \cite{Aoyama},
for the year 1998, we translate the income-tax
$t$ to the income $x$ as $t=0.3 x$ in the region above 50 million yen.
On the other hand, for the data sets of the income and the employment income,
we simply sum up these.
From this, we can gain the profile of the personal income distribution
over the 44 years.
However, we have a point to notice.
From the year 1965, all persons with income
greater than 20 million yen must file their final declaration to the JTA
under the Japanese tax system.
Hence individuals with employment income greater
than 20 million yen must file their final
declaration individually. Thus they are included in the income data.
For this reason, we use only the income data for the range greater than
20 million yen.
Although there are overlapping persons in the range less
than 20 million yen,
no detailed information to remove this ambiguity is possible.
We call the existence of the overlapping person as the
{\itshape overlapping effect}.
The effect of this will be stated later.
The example of above process for the year 1998 is summarized in the fourth
column in Table.
From these process we have the data for 51,493,254 individuals;
about 80\% of all workers in Japan.

\begin{table}[tb]
\centerline{Table: Japanese personal income distribution in the year 1998.}
\centerline{
\begin{tabular}{|c||r|r||r|}\hline
Classification &\multicolumn{3}{c|}{Number}\\\cline{2-4}
$x$-million yen&\multicolumn{1}{c|}{Income}
&\multicolumn{1}{c||}{Employment}
&\multicolumn{1}{c|}{Total}\\\hline\hline
$x\leq 0.7$  &14,496   &3,294,000 &3,381,848  \\\cline{1-2}
$0.7<x\leq 1$&73,352   &          &           \\\hline
$1<x\leq 1.5$&359,157  &4,639,000 &5,539,896  \\\cline{1-2}
$1.5<x\leq 2$&541,739  &          &           \\\hline
$2<x\leq 2.5$&565,835  &6,783,000 &7,933,824  \\\cline{1-2}
$2.5<x\leq 3$&584,989  &          &           \\\hline
$3<x\leq 4$  &954,901  &8,118,000 &9,072,901  \\\hline
$4<x\leq 5$  &687,057  &6,587,000 &7,274,057  \\\hline
$5<x\leq 6$  &497,438  &4,796,000 &5,293,438  \\\hline
$6<x\leq 7$  &375,485  &3,485,000 &3,860,485  \\\hline
$7<x\leq 8$  &288,141  &2,428,000 &2,716,141  \\\hline
$8<x\leq 9$  &379,716  &1,647,000 &3,129,716  \\\cline{1-1}\cline{3-3}
$9<x\leq 10$ &         &1,103,000 &           \\\hline
$10<x\leq 12$&229,205  &1,995,000 &2,439,917  \\\cline{1-2}
$12<x\leq 15$&215,712  &          &           \\\hline
$15<x\leq 20$&190,524  &394,000   &584,526    \\\hline\hline
$20<x\leq 25$&140,533  &79,000    &140,533    \\\cline{1-1}\cline{3-3}
$25<x\leq 30$&         &98,000    &           \\\cline{1-2}\cline{4-4}
$30<x\leq 50$&82,519   &          &82,519     \\\cline{1-2}\cline{4-4}
$50<x$       &43,455   &          &43,455     
\\\hline\hline
Total     &6,224,254&45,446,000&51,493,254 \\\hline
\end{tabular}
}
\end{table}

\begin{figure}[tb]
 \begin{center}
  \scalebox{.8}{\includegraphics{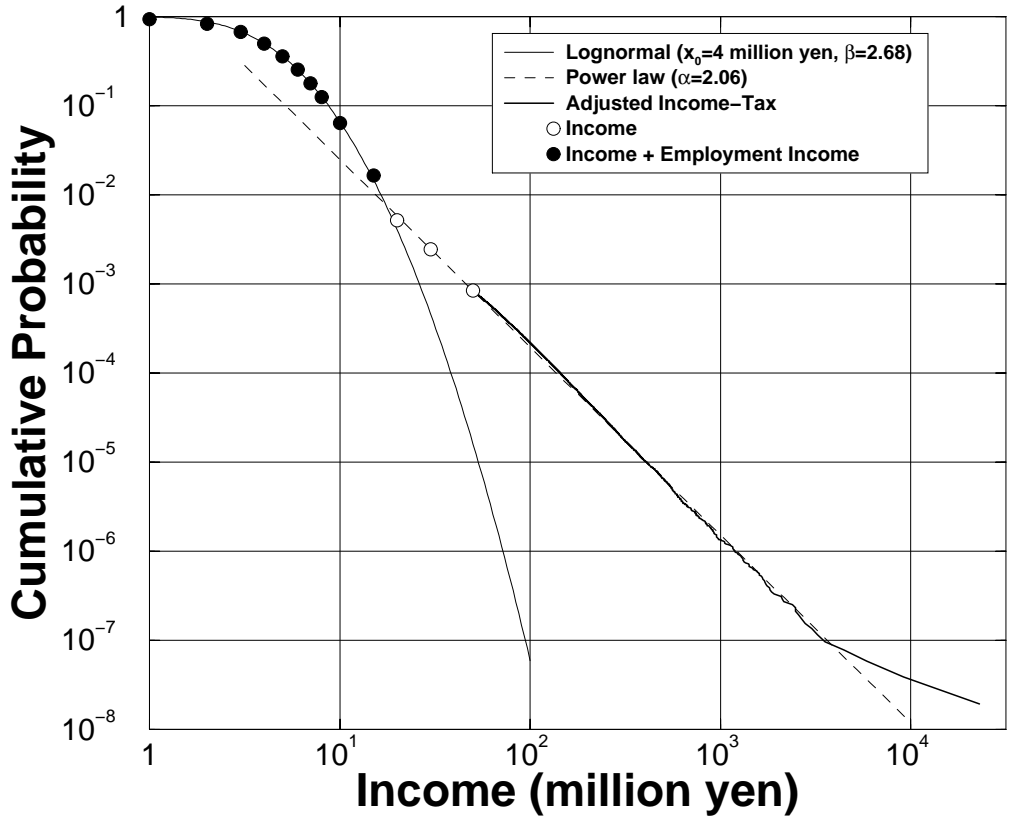}}
 \end{center}
 \caption{The cumulative probability
of the Japanese personal income in the year 1998.}
\label{fig1}
\end{figure}

The distribution for the year 1998 is shown in Fig.~\ref{fig1}.
In this figure, we take the horizontal axis as the logarithm of the income
with the unit of million yen,
and the vertical axis as the logarithm of the cumulative probability $P(x\leq)$.
The cumulative probability is the probability finding the person
with the income greater than or equals to $x$.
In the continuous notation it is defined by $P(x\leq)=\int_{x}^\infty dy p(y)$.
In this figure the bold solid line corresponds to the adjusted income-tax data.
The open circle emerges form only the income data.
The filled circle is the result of the sum of the income data and the
employment income data.
The dashed line and the thin solid line are the fitting of the linear
and the lognormal functions respectively.
We recognize from this figure that the top $1\%$ of the distribution
follows Pareto law with $\alpha=2.06$.
On the other hand $99\%$ of the distribution follows the
lognormal distribution. The mean value of the lognormal distribution
$x_0$ is about 4 million yen, and Gibrat index is $\beta=2.68$.
The change from the lognormal distribution to the power law distribution
does not occur smoothly. It is considered that the origin consists in
the {\itshape overlapping effect} stated before.
This is because, if we use only the employment income data for the
range less than 20 million yen to maximally remove the overlapping person,
the number of person in each classifications
in that region decrease. Hence the cumulative probability of each income
becomes small. This effect is expressed as the downward movement of
the filled circle in Fig.~\ref{fig1}.
Although the value of $x_0$ and $\beta$ changes, the pattern of
the lognormal distribution is not modified.
This is because the middle income distribution of employment income data
follows the lognormal distribution with $x_0=4$ million yen and $\beta=2.75$.

Although the detailed data for the high income is only possible for the
year 1998, the overall profile of the
distribution is available from the income data and the employment income data
as recognized from Fig.~\ref{fig1}.
Moreover as seen from Fig.~\ref{fig1}, the value of $\alpha$,
the absolute value of the slope of the high income distribution, is
available from only the income data, open circles in Fig.~\ref{fig1}.  
Hence the reliable value of indexes specifying the distribution is
calculable over the 44 years.

\begin{figure}[tb]
 \begin{center}
  \scalebox{.45}{\includegraphics{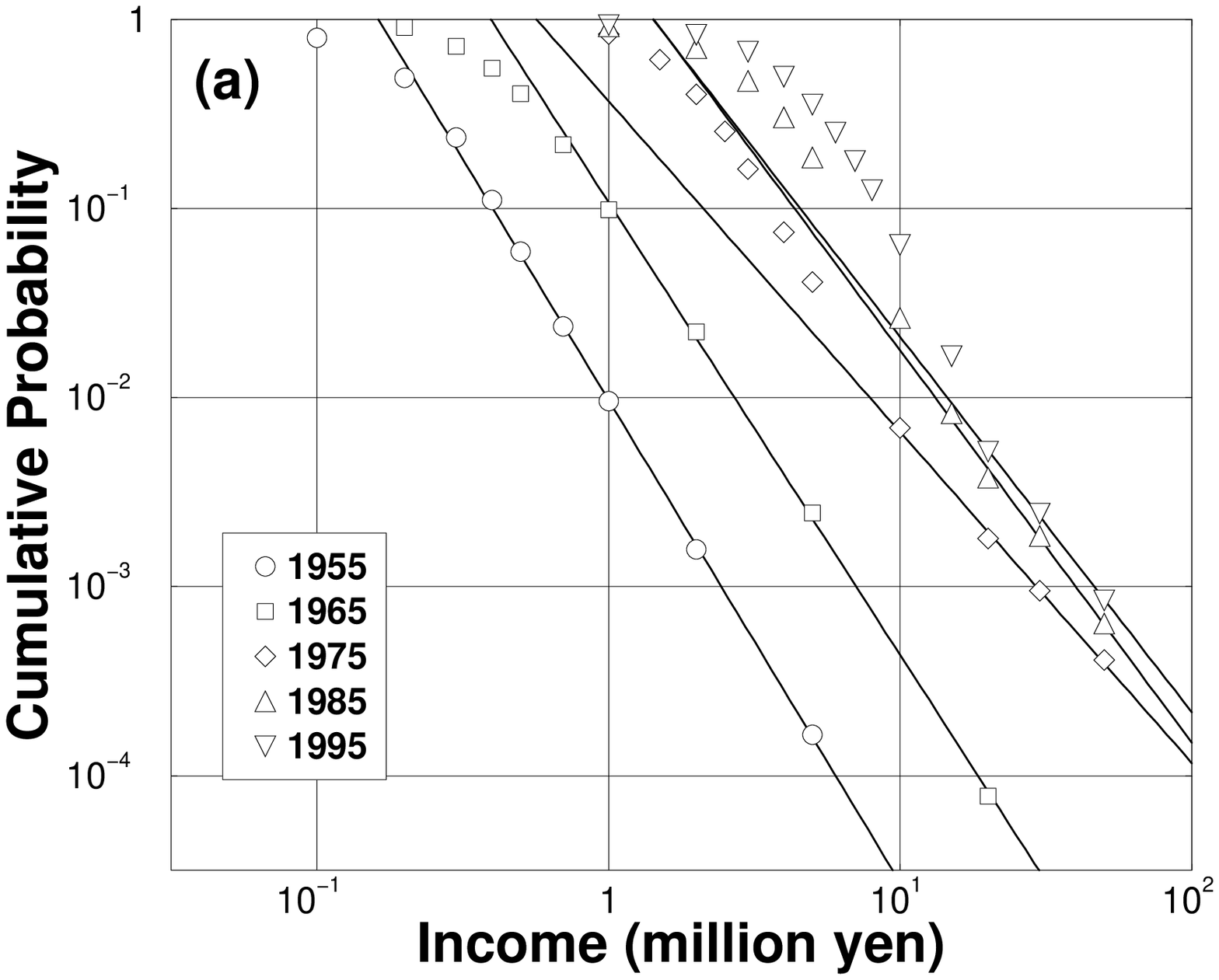}}
  \scalebox{.45}{\includegraphics{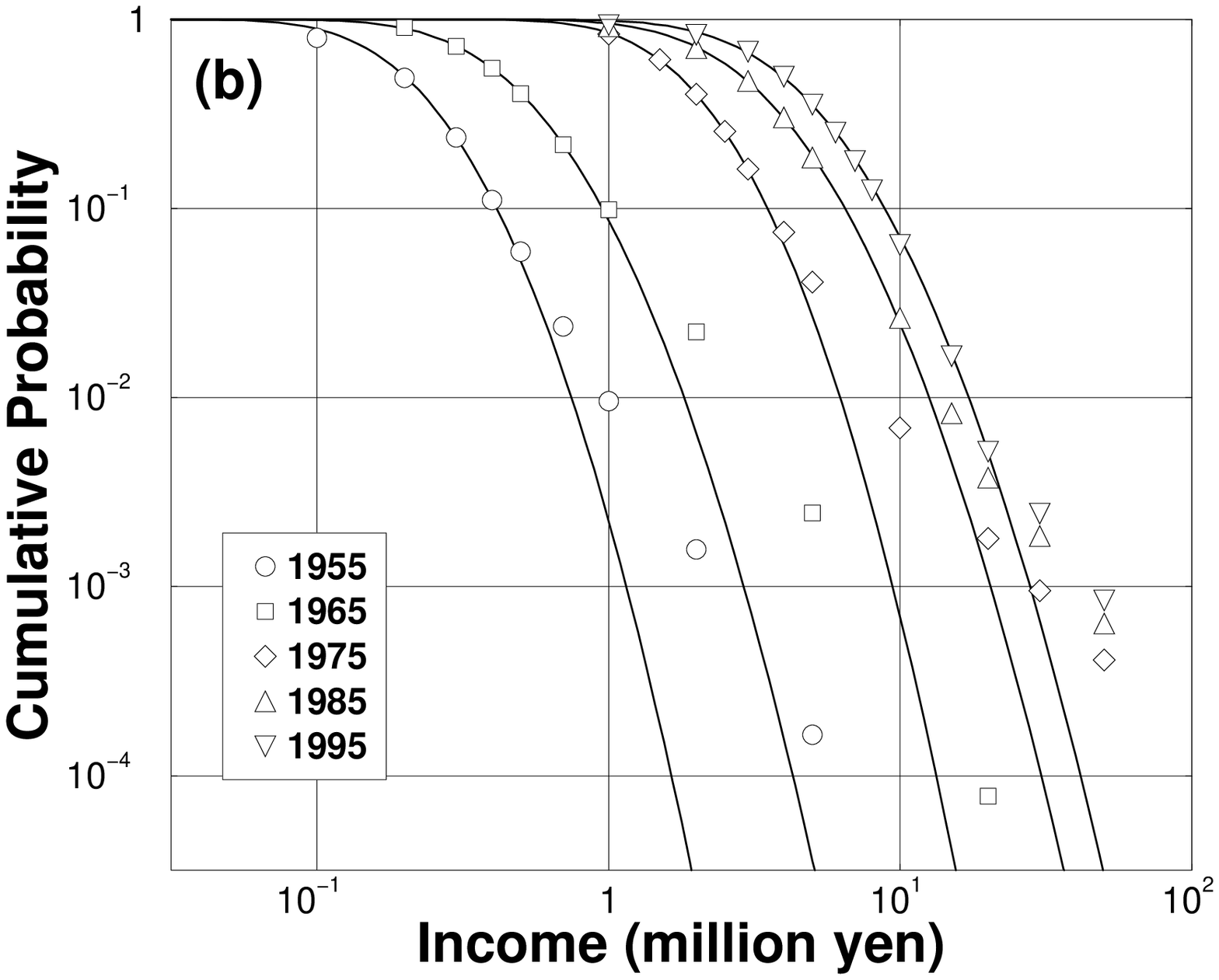}}
 \end{center}
 \caption{The cumulative probability of the Japanese
personal income for some years.}
 \label{fig2}
\end{figure}

The example of the distribution for some years are shown in
Fig.~\ref{fig2}~(a) and (b).
In these figures the plotted data sets are same for each figures.
The solid lines in Fig.~\ref{fig2}~(a) are the linear
fitting of the data for the high income.
We recognize from this figure that less than top 10\% of the income
is well fitted by Pareto law.
However the slope of each lines differs from each other.
Hence Pareto index differs from year to year.
In Fig.~\ref{fig2}~(b), the solid lines are the lognormal fitting.
It is confirmed from this figure that greater than 90\% of
persons follow the lognormal distribution. However the curvature of each
lines differs from each other.
Thus Gibrat index, the variance of the lognormal distribution,
differs from year to year.
The movement of the distribution toward the high income consists
in the increase of the mean income.
Hence this change is characterized by the change of 
the mean value of the lognormal distribution.
If we normalize the income by the
inflation or deflation rate, this movement may be deamplified.
However, if this manipulation is applied, the profile of the
distribution is not modified.

From the analysis of the data sets over the 44 years 1955-98, we can confirm
that the distribution pattern of the personal income is expressed
as the lognormal with power law tail. This distribution pattern
coincides with the result of
Ref.~\cite{lognormal}.
Though the analysis of Ref.~\cite{lognormal} and this article are for different
years and different countries, the same distribution pattern is observed.
Hence we can expect that the lognormal distribution with power law tail
of the personal income is a universal structure.
However
the indexes specifying the distribution
differ from year to year as recognized from Figs.~\ref{fig1} and \ref{fig2}. 
Thus we study the temporal change of these indexes in next section.

\section{The temporal change of the distribution}

\begin{figure}[tb]
 \begin{center}
  \scalebox{.45}{\includegraphics{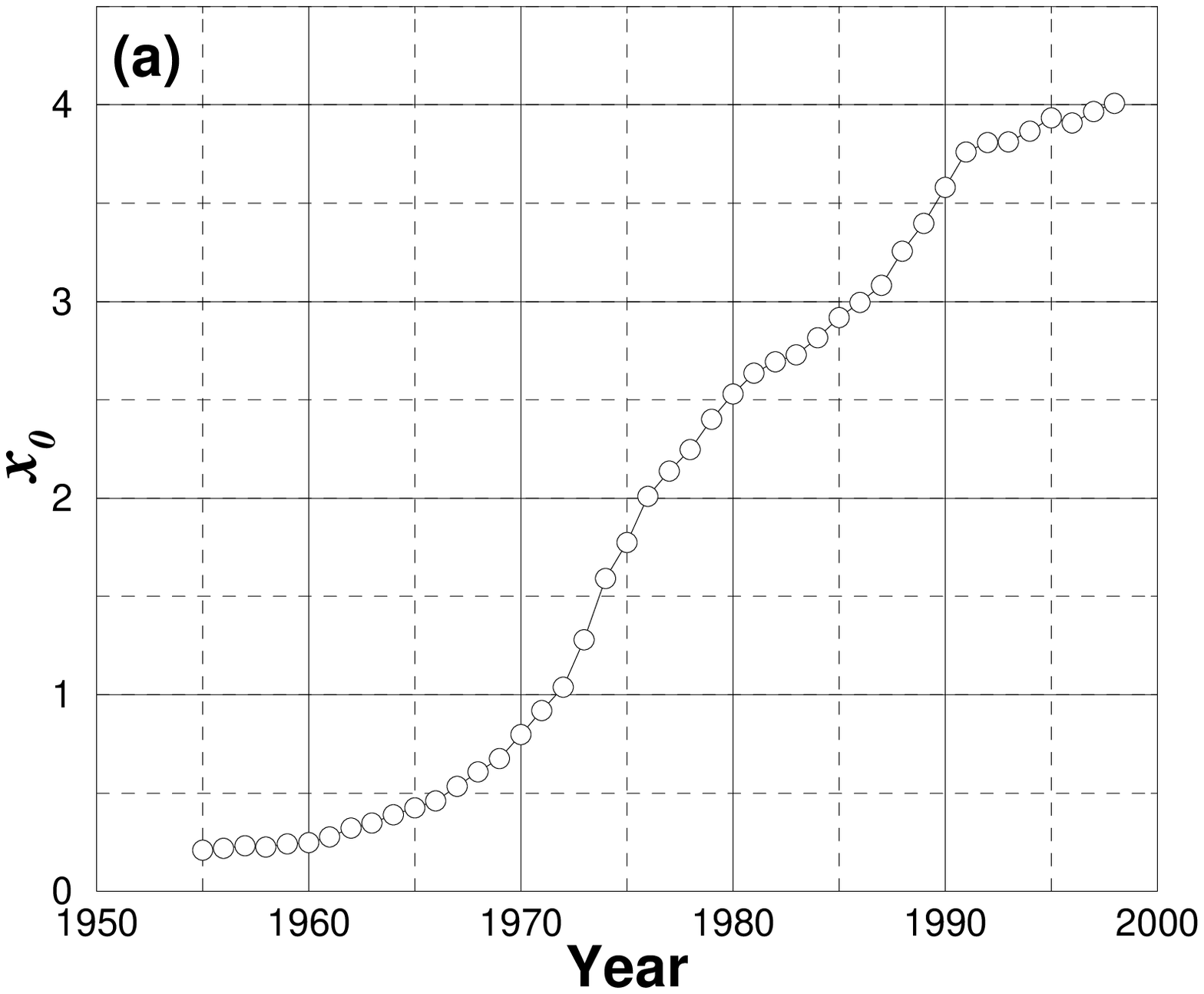}}
  \scalebox{.45}{\includegraphics{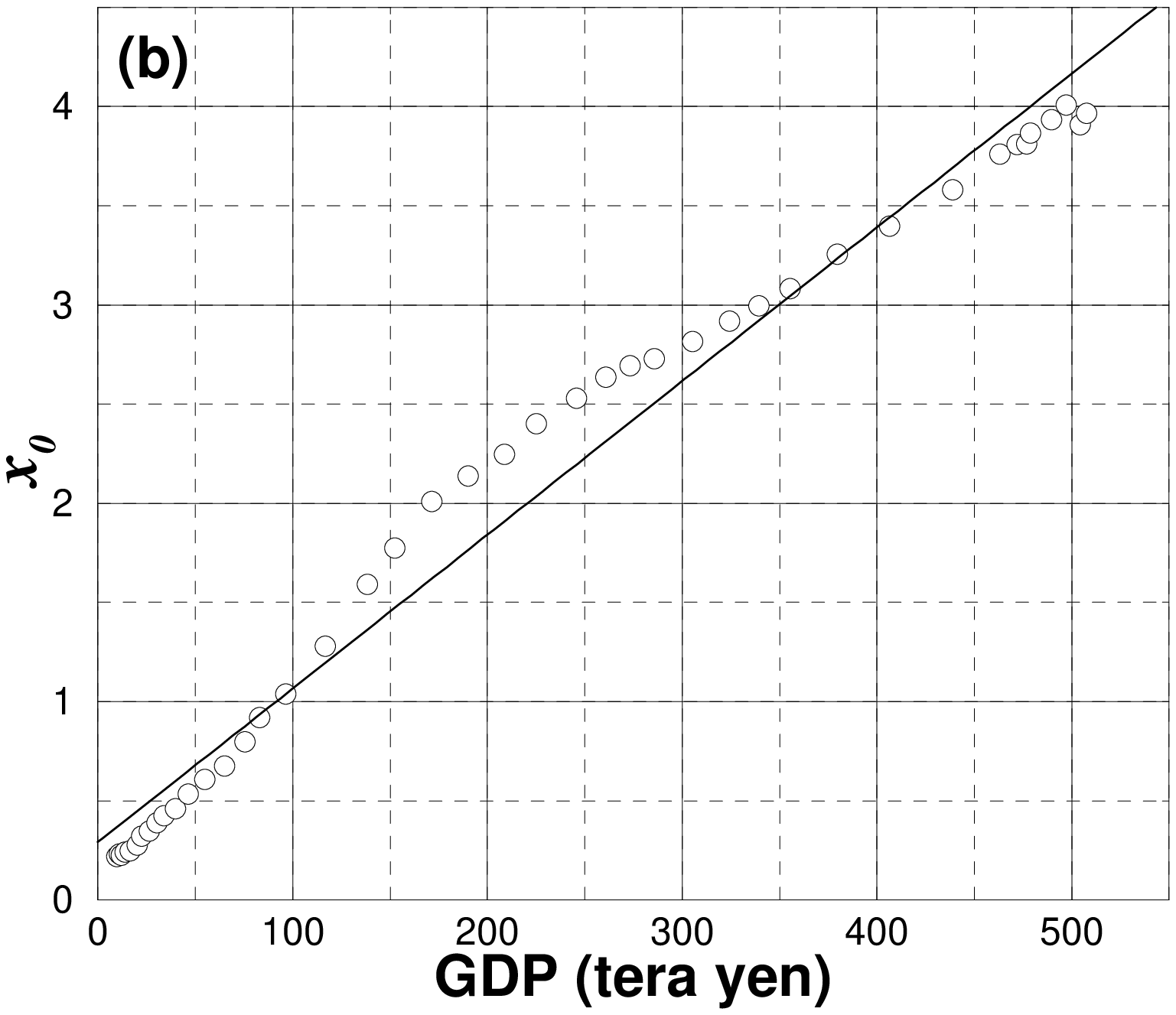}}
 \end{center}
\vspace{-5mm}
 \caption{The temporal change of $x_0$ (a), and the correlation
 between $x_0$ and the GDP (b) over the 44 years 1955-98.}
 \label{fig3}
\vspace{0mm}
 \begin{center}
  \scalebox{.45}{\includegraphics{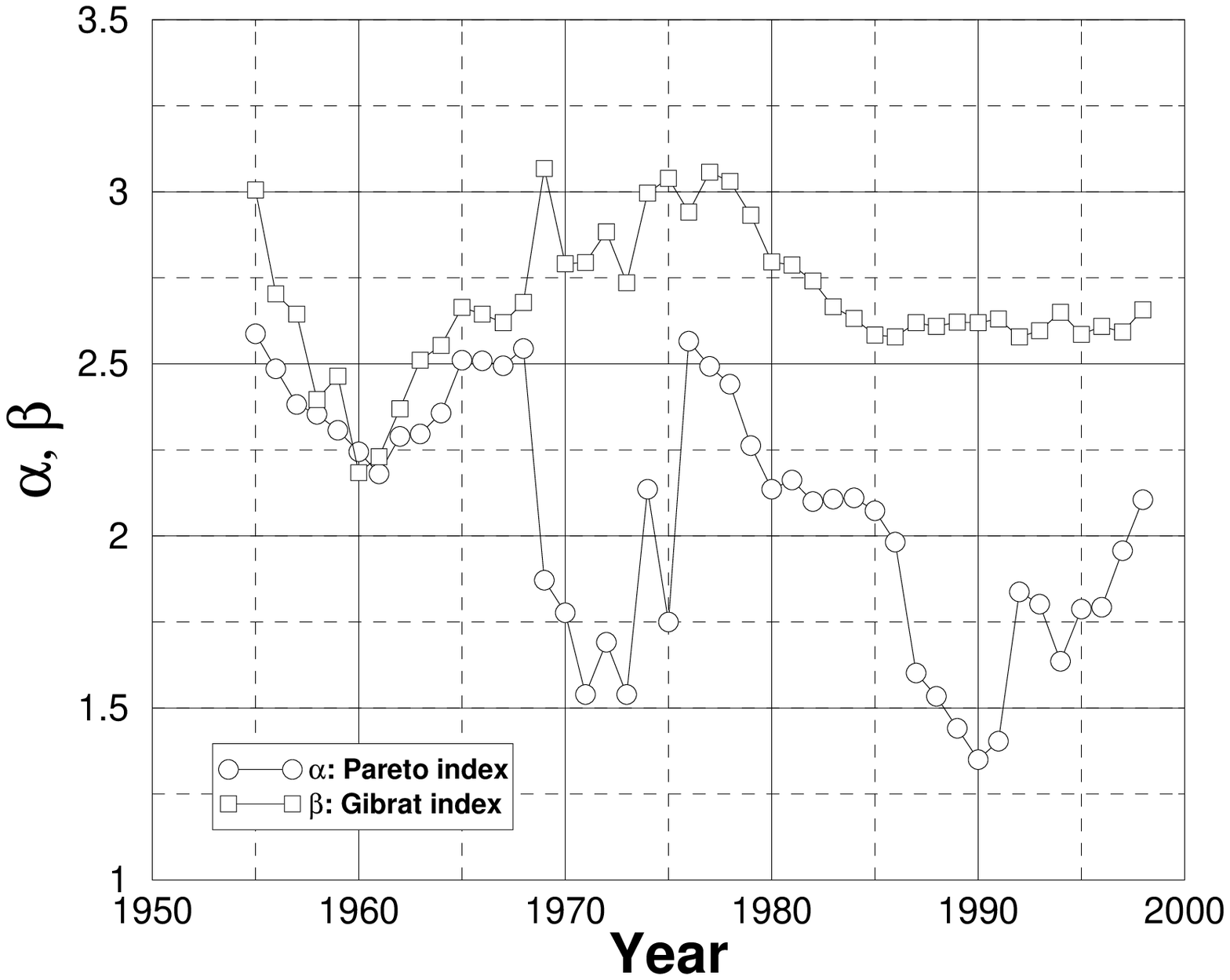}}
 \end{center}
\vspace{-5mm}
 \caption{The change of $\alpha$
 and $\beta$ over the 44 years 1955-98.}
\label{fig4}
\end{figure}

Firstly we consider the change of $x_0$ of the lognormal distribution.
As same as previous section, we can gain the change of it
over the 44 years 1955-98 by the numerical fitting.
The result is shown in Fig.~\ref{fig3}~(a).
In this figure the horizontal axis is the year and the vertical
axis is the value of $x_0$ with
the unit of million yen.
We recognize from this figure that the personal income rapidly
increases from the end of the 60s, and this condition 
continued until the beginning of the 90s.
It is reasonable to assume that the origin of the change of $x_0$ 
consists in the growth of the country.
To confirm this assumption we consider the correlation between $x_0$ and 
the Gross Domestic Product (GDP), and gain Fig.~\ref{fig3}~(b).
In this figure the horizontal axis is the GDP with the unit of tera yen
and the vertical axis is the value of $x_0$ with
the unit of million yen.
The solid line in this figure is
$
x_0\propto 0.00775G,
$
where $G$ is the GDP with the unit of tera yen.
Hence $x_0$ and the GDP linearly correlate.

Secondly, we consider the change of $\alpha$ and $\beta$ over the 44 years.
We have Fig.~\ref{fig4} from the numerical fitting of the distribution.
In this figure the horizontal
axis is the year and the vertical axis
is the value of $\alpha$ and $\beta$.
The open circle and square correspond to $\alpha$ and
$\beta$ respectively.
It is recognized from this figure that these
indexes correlate with each other around the year 1960 and 1980.
However these quantities
have no correlation in the beginning of the 70s and after the year 1985.
In the range where $\alpha$ and $\beta$ change independently,
the strongly changing index is $\alpha$.
Especially after the year 1985, $\beta$ stays almost same value.
This means that the variance of the middle income is not changing.
From these behaviors of $\alpha$ and $\beta$, it is considered
that there are some factors causing
no correlation between
$\alpha$ and $\beta$, and mainly effecting to $\alpha$.

\begin{figure}[tb]
 \begin{center}
  \scalebox{.45}{\includegraphics{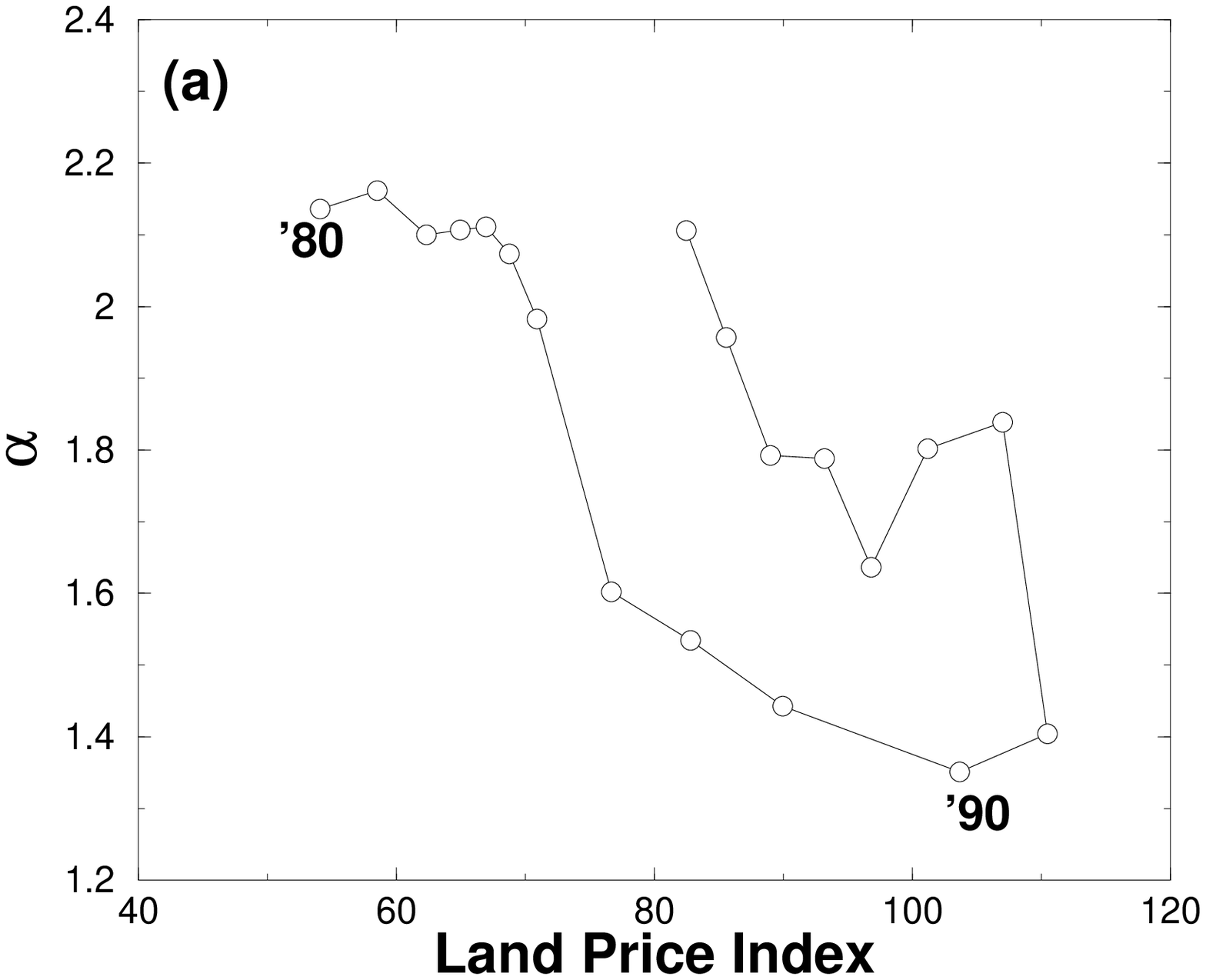}}
  \scalebox{.45}{\includegraphics{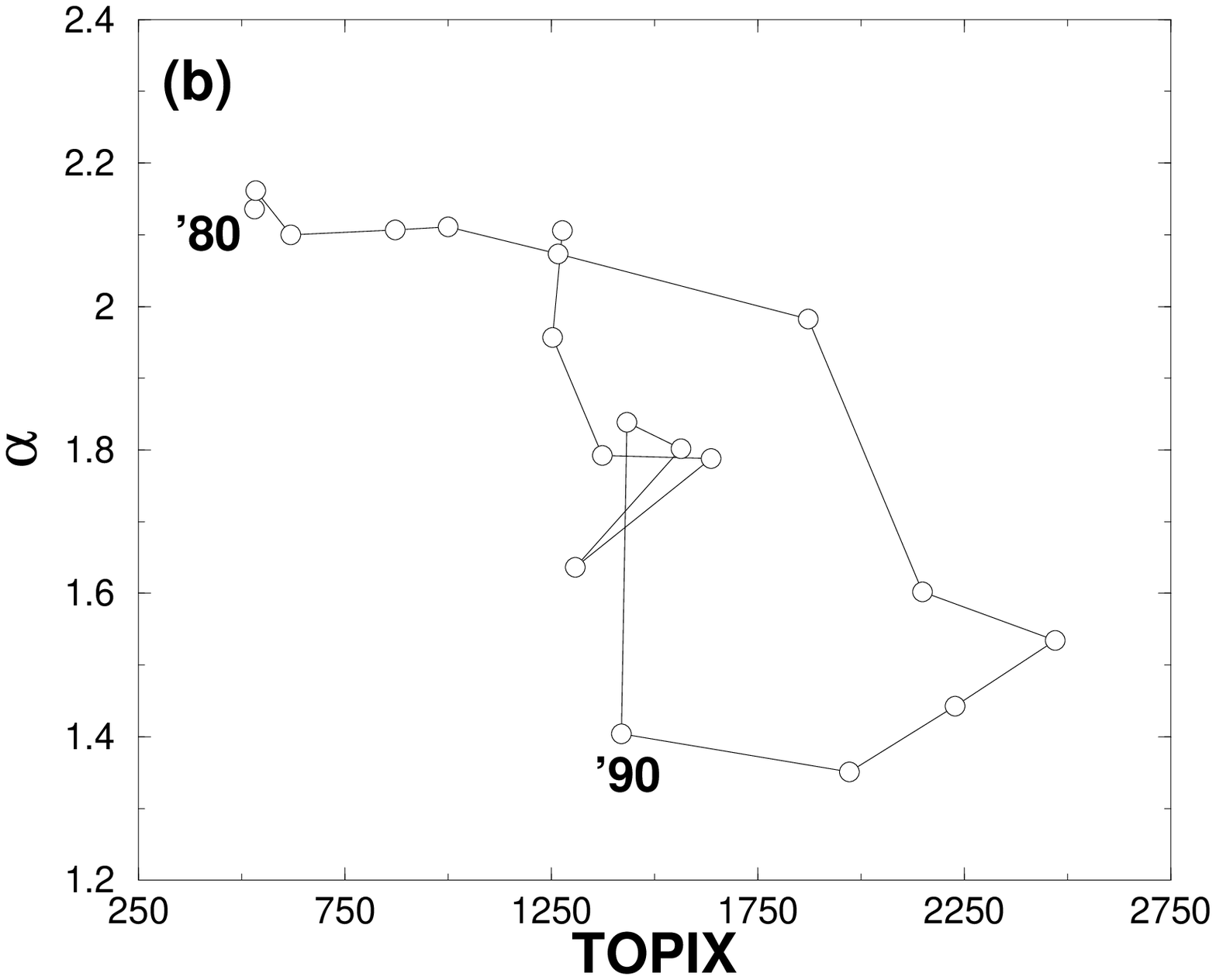}}
 \end{center}
\vspace{-5mm}
 \caption{The correlation between $\alpha$ and the land price
 index (a) and the TOPIX (b).}
 \label{fig5}
\vspace{3mm}
 \begin{center}
  \scalebox{.8}{\includegraphics{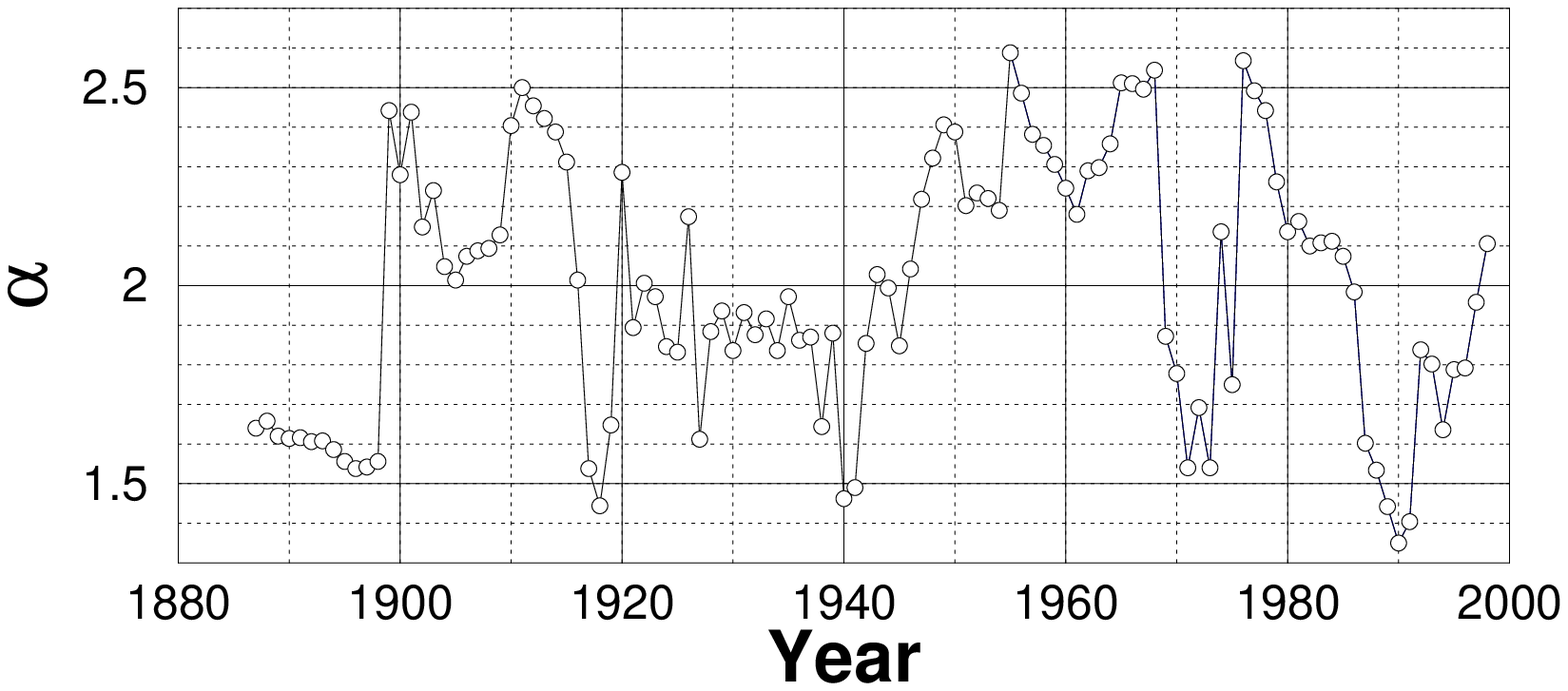}}
 \end{center}
\vspace{-5mm}
 \caption{The change of $\alpha$ over the 112 years 1887-1998.}
 \label{fig6}
\end{figure}

Thirdly to study the origin of the change of $\alpha$ more detail, we consider
the correlation between $\alpha$ and the asset price
such as the land price and the stock price
over the recent 19 years 1980-98.
The correlation between the land price index and $\alpha$ is shown in
Fig.~\ref{fig5}~(a).
The land price index is defined as the mean value of the
land price of that year, normalized so that the
land price as of May 31, 1990 is 100.
In this figure the horizontal axis is the land price index and the
vertical axis is the value of $\alpha$.
We recognize from this figure that these two indexes linearly
anti-correlate.
If the land price increases, the value of $\alpha$ becomes small.
The correlation between $\alpha$ and the TOPIX is shown in Fig.~\ref{fig5}~(b).
In this figure the horizontal axis is the TOPIX and the
vertical axis is the value of $\alpha$.
Though the correlation is not so
strong in comparison with the case of the land price index,
the anti-correlation is observed.
If the TOPIX increases, the value of $\alpha$ becomes small.
Hence the change of the asset price is
one of the origin of the change of Pareto index.

Lastly,
as mentioned previously, $\alpha$
is mainly derived from the income data.
Hence the change of $\alpha$ can be analyzed over the 112 years 1887-1998.
The data analysis are shown in Fig.~\ref{fig6}.
In this figure the horizontal axis is the year and the
vertical axis is the value of $\alpha$.
The mean value of Pareto index is $\bar\alpha=2$, and $\alpha$
fluctuates around it.
Japan has experienced various
business conditions except for the socialist economy
over the 112
years.
Hence it is expected that this behavior of Pareto index is 
applicable for another countries.
However to confirm this conjecture is a future problem.
The mean value of Pareto index $\bar\alpha=2$ in this case
is worth to compare with the case of the Japanese
company size and that of the debts of bankrupt companies
\cite{Aoyama},\cite{Katori}.
In these cases the distribution follows the power law with $\alpha=1$;
Zipf law. 
\vfill
\section{Summary and discussion}
In this article we investigated
the Japanese personal income distribution
in the high income range over the 112 years 1887-1998, and
that in the middle income range over the 44 years 1955-98.
We confirmed from data analysis that
the personal income follows
the lognormal distribution with power law tail.
This structure have been observed in the analysis for the different years and
countries.
Hence it is the universal structure that 
the personal income distribution takes the form of the
lognormal with power law tail.

From the analysis of the change of $\alpha$ and $\beta$, we confirmed that
these quantities should not
necessary correlate each other.
This means that different mechanisms are working in the distribution of
the high income range and that of the middle income range.
One of the origin of no correlation is the change of the
asset price such as the land price and the stock price.
The high income distribution is strongly affected
by the asset price.

From the analysis in this article, we propose that the following
conditions must be
satisfied by mathematical models.
\begin{enumerate}
\item The model must express the lognormal distribution with
power law tail.
\item There are some factors causing no correlation between
$\alpha$ and $\beta$, and mainly effecting to $\alpha$.
\item Pareto index must fluctuate around $\bar\alpha=2$.
\item Gibrat index fluctuates contradictory to the random walk. 
\end{enumerate}

The most simple model intended to express the mechanism of the
distribution  is given by R.~Gibrat \cite{Gibrat} as the
multiplicative stochastic process defined by
$x(t+1)=b(t)x(t)$. Here $x(t)$ is the income at the time $t$ and
$b(t)$ is the ideally independently distributed
stochastic variables with positive values.
Hence if this process is iteratively used, we have
$x(t+1)=b(t)\cdot b(t-1)\cdots x(0)$.
Thus the logarithm of this derives
$\log x(t+1)=\log b(t) +\log b(t-1) +\cdots+\log x(0)$.
Hence $\log x(t+1)$ follows the normal distribution by
the central limit theorem.
Therefore $x(t+1)$ takes the form of the lognormal distribution.
However this process derives the monotonically increasing variance.
Moreover the power law tail can not be explained by this
process only, and
the simple multiplicative stochastic process
can not satisfy above conditions.
In present, there are some models deriving the power law tail \cite{newmodels}.
Though these models does not intend to explain the
personal income distribution, to 
investigate the applicability of these
models to the personal income is an interesting problem.

\vspace{10mm}
\noindent {\bfseries\Large Acknowledgments}

\vspace{5mm}
\noindent The author would like to thank H.~Aoyama (Kyoto Univ.)
and H.~Takayasu (Sony CSL) 
for useful discussions and advice, and
Mr.~Nakano (JTA) for conversation on the properties of the data
sets.
The author would also like to thank
M.~Katori (Chuo Univ.), M.~Takayasu (Hakodate Future Univ.) and
T.~Mizuno (Chuo Univ.)
for useful comments.

\end{document}